# Analysing the form of the confined uniaxial compression curve of various soils

Anh-Minh Tang [1], Yu-Jun Cui [1], Javad Eslami [1] and Pauline Défossez [2]

[1] Ecole des Ponts - ParisTech, UR Navier/CERMES, 6 et 8, avenue Blaise Pascal, Cité Descartes, Champs-sur-Marne, 77455 Marne La Vallée cedex 2, France.
[2] INRA, UR1158, Unité d'Agronomie rue Fernand Christ, 02007 Laon Cedex, France.

**Corresponding author**

Anh-Minh Tang

Ecole des Ponts – ParisTech
U.R. Navier/CERMES
6-8 av. Blaise Pascal, Cité Descartes
77455 Marne –la–Vallée
FRANCE
Tel: 33 1 64 15 35 63
Fax : 33 1 64 15 35 62
E-mail : tang@cermes.enpc.fr

**Abstract**

The soil compaction by vehicles is a major factor responsible for physical degradation of cultivated soils. Uniaxial confined compression tests are usually performed to characterise the compaction properties of soil. Two main forms of compression curve have been observed: (i) the bi-linear curve having an elastic rebound curve at low stresses and a linear virgin compression curve at higher stresses; (ii) the S-shaped curve having deviation of the virgin compression curve at high stresses. In the present work, uniaxial confined compression tests were performed on four soils having various textures and different plasticity. Tests were performed on undisturbed and remould samples, at various initial dry bulk densities and water contents. The S-shaped compression curves were observed more frequently when the clay content and/or the initial water content were high. In addition, the S-shaped curves were observed more frequently on remould soils than on undisturbed soils. The difference between the compression of air-filled pores and that of meso-pores storing water subjected to high capillary forces could explain the observed S-shaped curves.

## 1. Introduction

Soil compaction is one of the most important factors responsible for soil physical degradation (Pagliai et al., 2003). It alters the structure of cultivated soil and thus changes a number of key soil properties for crop production and the environment (Soane and van Ouwerkerk, 1994; Chan et al., 2006). Excessive soil compaction can have adverse effects on seedling emergence and plant rooting. It causes a decrease in oxygen diffusion so that denitrification can occur (Czyz, 2004). It can also increase the runoff risk due to the decrease in water infiltration. Obviously, it is important to control the mechanical impacts of agricultural machinery on soil structure to reduce excessive soil compaction (Alakukku et al., 2003). Soane and van Ouwerkerk (1994) concluded that the intensity of soil compaction is mainly due to two types of factors: (i) the applied load which depends on vehicle characteristics (axle load, tyre dimensions, tyre inflation pressure, and vehicle velocity); (ii) the soil mechanical strength which depends on soil intrinsic characteristics such as its texture, on soil quasi-permanent characteristics such as its carbon content and on short-term changing characteristics such as soil moisture condition and packing state.

Confined uniaxial compression test is usually performed to characterise the soil compressibility properties. The stress-strain relationship of soil obtained from this test is often described by two types of models: (i) elasto-plastic models; (ii) S-shaped models. According to the elasto-plastic models (bi-linear form), the compression curve (in this paper: void ratio, $e$, versus the logarithm of vertical stress, $\log\sigma_v$) can be divided into two parts: an elastic rebound part at low stresses and a linear virgin compression part at higher stresses (Fig. 1$a$). The transition point between the elastic rebound part and the virgin compression part is known as the soil precompression stress. The elasto-plastic models are the mostly used method to interpret the results obtained from uniaxial compression tests (Pereira et al., 2007; Mosaddeghi et al., 2007; Défossez and Richard, 2002; among others). As these models consider a constant slope for the virgin compression part, they can not satisfactorily describe the compression behaviour at high stresses. Indeed, experimental results generally show that the slope of virgin compression part decreases beyond a certain stress. In addition, when determining the precompression stress $\sigma_p$ using various methods (Dawidowski and Koolen, 1994; Arvidsson and Keller, 2004; Imhoff et al., 2004), significant error could be induced if the compression curve deviates from the linear trend at high stresses. In the work of Keller et al. (2004) the data at the highest stress levels were not used for $\sigma_p$ determination. This justifies the use of the S-shaped models that allows the compression curve to deviate at high stresses (Fig. 1$b$). Polynomial or sigmoidal functions are often adopted to fit the S-shaped compression curves (Baumgartl and Kock, 2004; Gregory et al., 2006).

Following the soil compaction processes described by Horn et al. (1995), soils are three-phase systems which undergo changes as soon as the external stresses exceed the internal soil strength. The decrease of soil volume upon stress increase involves firstly the air-filled pores. The pore-water pressure and the pore-air pressure are then increased and reduced with time due to the dissipation process. This process depends strongly on the permeability of the soil sample which also decreases during the compaction. In addition, hydraulic conductivity of the soil depends on the soil pore structure (Dexter et al., 2004; Horn, 2003). Thus the compression of cultivated soils corresponds to a complex process because many factors are involved. The present work attempts to analyse experimentally the parameters that control the form of the compression curve obtained from the confined uniaxial test (water content, dry bulk density, soil structure, soil plasticity).

## 2. *Materials and methods*

The soils were taken from four sites in France: (A) le Breuil, (B) Mons, (C) Avignon and (D) Epernay. The first site (A) was the « Breuil-Chenue » experimental forest site located in the Morvan (47°18'N, 4°4'E, centre of France) and described in Moukoumi et al. (2006). The soil was a sandy clay (Dystric Cambisol) where monospecific plantation plot have been conducted since thirty years (oaks, beech, spruce and Douglas fir). The second site (B) was the long-term field trial at the INRA centre of Estrée-Mons (50°N, 3°E, northern France) which involves a loamy soil (Hortic luvisol) described in Boizard et al. (2002). The third site (C) was a sugar beet field at the Avignon research centre of INRA (43°55'N, 4°53'E, south of France) (Chanzy et al., 1998). The soil was calcareous with a silty clay loam texture (Calcaric Cambisol). The fourth site (D) was an experimental site managed by the "CIVC" technical institute for Champagne wines. It is located in Epernay (49°N, 3°56'E, east of France) and the soil was calcareous with a clay texture (Calcaric Cambisol). Samples were collected on a plot where vine was planted in 1988.

The soil properties were determined following the French Standard for Geotechnical Engineering: the particle density was determined using water pycnometer on the soil sieved at 2 mm; the Atterberg limits were determined on the soil sieved at 0.4 mm; and blue value was determined using the Methylene blue absorption method on the soil sieved at 0.5 mm. The organic carbon content is 82.8 g/kg for soil A, 8.5 g/kg for soil B, 10.2 g/kg for soil C and 16.8 g/kg for soil D. Soil texture was classified following FAO-UNESCO (1974) system (after Jones et al., 2003) and USDA classification that are based on the particle size distribution. According to FAO classification, the texture of the tested soils varies from medium to fine. Some physical and chemical properties of the studied soils are presented in Table 1.

Confined uniaxial compression tests were performed on the soils taken from two layers (0 – 30 cm and 30 – 60 cm depth) at each site. Remould samples were tested for the topsoil layer (0 – 30 cm depth) which is frequently tilled whereas undisturbed samples were tested for the subsoil (30 – 60 cm) where compaction is persistent. For the preparation of remould samples, the topsoil was air-dried, grounded and sieved at 2 mm. Then the soil was wetted by spraying distilled water to achieve the desired water content and then stocked in a hermetically-sealed box for 24 h for water homogenization. Finally, the soil was poured directly into the Casagrande oedometer cell and compacted to achieve the desired initial bulk density. The apparatus (Casagrande oedometer cell) consists of an oedometer cell that can be installed in a loading frame so that a predetermined vertical stress can be applied to the specimen. The soil specimen is restrained laterally by a steel oedometer cell. The top and the bottom surfaces of the specimen are in contact with porous stone discs. The test is considered to be one dimensional as both specimen deformation and drainage only occur in the vertical direction.

For the preparation of undisturbed samples, a core sampler of 70-mm high, 150-mm in inner diameter and 1-mm thick, was pushed vertically into the soil. The soil cylinders were then wetted by spraying distilled water to achieve the desired water content. They were protected by plastic film during 24 h for water homogenization. Mosaddeghi et al. (2007) spent 2 days to saturate their soil cylinders of 50-mm high and 98.6 mm in diameter, from the bottom only. In the present work, the soil cylinders were wetted from both the bottom and the top; the duration of 24 h was considered long enough to reach water homogenization. Then, the soil specimen (70 mm in diameter and 20 mm high) was trimmed directly from the cylinder and inserted into the oedometer cell. When a drying process was involved (initial water content lower than the natural one), the soil cylinder was air-dried for 2 h, and then covered by plastic film for 6 h. The procedure was repeated until the desired water content was achieved. This drying process prevented the soil from cracking during the preparation. Finally, as for the wetted samples, the soil specimen was prepared by trimming and inserted into the oedometer cell.

After installation of the soil specimen, a low vertical stress (15 kPa) was applied to ensure a good contact between the piston and the soil surface. Vertical stresses of 30, 50, 100, 200, 300, 600, 800 kPa were applied step-by-step. Soil specimen was unloaded following the same steps until the initial stress (15 kPa). Each stress was applied for 5 min and the vertical displacement was recorded at the end of each step with an accuracy of ±0.001 mm. Note that in the studies of the compaction behaviour of agricultural soils, the load is often applied quickly (5 min to 45 min) to simulate the short duration loading by a tractor vehicle in the field, about 0.5 s (after Keller et al., 2004). At the end of each test, the soil sample was taken out of the cell and its dimension was taken using a calliper with an accuracy of ±0.001 mm. Its final water content was determined by oven-drying at 105 °C for 24 h. Thus, the final void ratio and the degree of saturation could be determined. The void ratio ($e$) corresponding to each vertical stress was back-calculated using the final void ratio and the measured displacement. The degree of saturation ($S_r$) and the volumetric air content ($V_a/V$) were calculated from the initial values as well as the measured displacement with the assumption that no water was squeezed out when $S_r$ was lower than 100%.

In total, 41 uniaxial compression tests were performed. The initial parameters (void ratio, $e_i$; water content, $w_i$; dry bulk density, $\rho_i$; degree of saturation, $S_{ri}$) and the final parameters (void ratio, $e_f$; water content, $w_f$; dry bulk density, $\rho_f$; degree of saturation, $S_{rf}$) are presented in Table 2 for each test. Tests on the remould specimens were performed on soils A, C, and D. For each soil, three values of $w_i$ were considered; for each value of $w_i$, the soil was compacted at three values of $\rho_i$ ranging from 0.98 to 1.44 Mg/m$^3$. Nine tests on remould specimens were thus performed for each soil. In the case of undisturbed specimens, three values of $w_i$ were considered for soils A, C, and D, and five values of $w_i$ were considered for soil B.

## 3. Results

The results of the remould specimens of soil A are presented in Fig. 2. Only the results obtained from the loading path (vertical stress increased step-by-step from 15 to 800 kPa) are presented. For each loading step, the ratio of the change of $e$ to the change of the logarithm of $\sigma_v$ (-d$e$/dlog$\sigma_v$) was calculated and plotted versus $\sigma_v$. The results from the other tests are presented using similar plots in Fig. 3 (soil C, remould), Fig. 4 (soil D, remould), Fig. 5 (soil A and soil B, undisturbed), and Fig. 6 (soil C and soil D, undisturbed).

In the case of the remould specimens of soil A (Fig. 2), the $e$-$\log\sigma_v$ plot shows that almost all the tests had the bi-linear form. Moreover, when analysing the plot -$de/d\log\sigma_v$ versus $\sigma_v$, the tests T1, T2, T3, T4, T5 and T6 showed an increase of -$de/d\log\sigma_v$ followed by a slight decrease. That corresponds to a lightly S-shaped compression curve that is not visible in the $e$-$\log\sigma_v$ plot. For the tests T7, T8 and T9, -$de/d\log\sigma_v$ increased continuously when $\sigma_v$ increased. For all the tests showed in Fig. 2, the initial degree of saturation $S_{ri}$ varied between 0.3 and 0.7. Upon compression, $S_r$ increased up to 0.5 – 0.9 and $V_a/V$ decreased from 0.15 – 0.40 to 0.05 – 0.20.

In the case of the remould specimens of soil C (Fig. 3), three groups of tests can be identified. The first one (T18, T21, and T24) had the bi-linear compression curve. The $S_r$ - $\log\sigma_v$ plot shows that this group had the lowest $S_{ri}$ ($S_{ri}$ = 0.3 – 0.5) and the final value of $S_r$ was equal to 0.7. This group had equally the highest initial values of $V_a/V$ (0.3 – 0.4) and $V_a/V$ decreased to 0.1 at 800 kPa vertical stress. For each test in this group, -$de/d\log\sigma_v$ was not changed significantly when the stresses were higher than 100 kPa; this means that the compression curve ($e$-$\log\sigma_v$ plot) was almost linear at high stresses. The second group (T19, T22 and T25) had a clear S-shaped compression curve: when $\sigma_v$ was increasing, -$de/d\log\sigma_v$ firstly increased and reached its maximum values (ranging from 0.3 to 0.6) for the loading step of 50 – 100 kPa. The value of -$de/d\log\sigma_v$ decreased to 0.2 at the maximum stress. In the third group (T20, T23 and T26), -$de/d\log\sigma_v$ remained almost constant. This group had the highest $S_{ri}$ (higher than 0.7) and the smallest initial air volumetric fraction $V_a/V$ (smaller than 0.15). During compression, the soil reached its saturation state at a stress higher than 100 kPa for the tests in the third group. For all the tests of the remould specimens of soil C, compression increased $S_r$ from 0.3 – 0.9 to 0.8 – 1.0 and decreased $V_a/V$ from 0.05 – 0.40 to 0 – 0.10. Note that $S_r$ and $V_a/V$ were calculated with the assumption that no water was squeezed out of the soil when $S_r$ < 1. When $S_r$ = 1, the soil water started to be squeezed out and the soil was considered to be saturated at higher stresses.

All the tests of the remould specimens of soil D (Fig. 4) had a S-shaped compression curve. For these tests, $S_{ri}$ ranged from 0.4 to 0.9 and $S_r$ increased to 0.9 – 1.0 at the maximum vertical stress. During compression, $V_a/V$ decreased from 0.05 – 0.40 to 0 – 0.05. In addition, -$de/d\log\sigma_v$ reached its maximum value under a vertical stress $\sigma_v$ comprised between 50 and 400 kPa; this corresponds to the inflection point in the compression curve ($e$ - $\log\sigma_v$ plot).

The compression tests on the undisturbed specimens of soil A and soil B are presented in Fig. 5. Only the test T11 showed a clear S-shaped compression curve; the other tests had the compression curve with bi-linear form. Note that the test T11 had the highest $e_i$ ($e_i$ = 1.4) and initial air volumetric fraction ($V_a/V$ = 0.4). Similar observations can be made from the compression tests on the undisturbed specimens of soil C and soil D (Fig. 6): almost all the tests had the bi-linear compression curve except the test having the highest $e_i$ and initial $V_a/V$ (T41).

The compression index $C_c$ was defined as the maximum value of -$de/d\log\sigma_v$ along the loading path. That corresponds to the slope at the inflection point on the compression curve. This method for $C_c$ determination is similar to that presented by Gregory et al. (2006) and Baumgartl and Köck (2004). The calculated compression index $C_c$ at various $\rho_i$ and $w_i$ are plotted in Fig. 7. For remould soils, the results are presented in three groups of $w_i$ for each soil. For all the soils, $C_c$ was smaller at bigger values of $\rho_i$. At a same value of $\rho_i$, $C_c$ of soil

A was not affected by $w_i$ (Fig. 7*a*). For soil C (Fig. 7*b*), $C_c$ at the highest $w_i$ (28.5%) was smaller than that at lower $w_i$ (17 and 22%). The effect of $w_i$ on $C_c$ at a same value of $\rho_i$ was clearer for soil D (Fig. 7*c*): the higher the initial water content $w_i$, the lower the compression index $C_c$. For the undisturbed soils, $C_c$ was plotted versus $w_i$ (Fig. 7*d*). No clear effect of $w_i$ on $C_c$ can be observed.

## *4. Discussion*

Several authors reported the S-shaped curves (Baumgartl and Kock, 2004; Gregory et al., 2006), but few analysed the factors that affect the shape of the compression curve. The results obtained in the present work enable this analysis. In the case of remould soils, the S-shaped curves were more frequently observed on the high plasticity soils (with high plasticity index, *PI*) than on the low plasticity soils (with low plasticity index, *PI*). For instance, almost all the tests on soil A (Sandy loam, *PI* = 7) presented the bi-linear compression curve and almost all the tests on soil D (Clay, *PI* = 20) had the S-shaped one. In the case of soil C (Silty clay loam, *PI* = 11), either bi-linear or S-shaped curves were observed. The results of the remould specimens of soil B allowed identifying the effect of the water content on the compression behaviour. Obviously, the three types of compression curves can be distinguished clearly when plotting the degree of saturation ($S_r$) or air volumetric fraction ($V_a/V$) versus vertical stress ($\sigma_v$), (see Fig. 3), and it is not the case when plotting void ratio (*e*) versus vertical stress ($\sigma_v$). This means that the initial water content, $w_i$ (or $S_{ri}$, $V_a/V$) has a stronger influence on the compression behaviour (i.e. the shape of the compression curve) than the initial void ratio ($e_i$).

Almost all the tests performed on the undisturbed soils had the bi-linear compression curve. Note that the remould specimens of soil C had the S-shaped compression curve when $S_{ri}$ was higher than 0.5, whereas the undisturbed specimens had a bi-linear compression curve even though when $S_{ri}$ was as high as 0.7 – 0.8. As a conclusion, when the structure of the soil was destroyed as for the remould samples, the S-shape could be expected.

Following Dexter et al. (2004), soil structure can be described by three pore sizes: micro-pore, meso-pore and macro-pore. The micro-pore corresponds to the matrix domain (intra-aggregates) with the range of 0.1 – 10 µm (after Kutilek et al., 2006). The meso-pore corresponds to the structural domain (inter-aggregates) and its size varies in the range of 10 – 50 µm. The macro-pores in agricultural soils are formed generally by roots, worms or cracking and its size is larger than 50 µm according to Wierman et al. (2000). These descriptions of pore sizes are applied in the present work when analysing the compaction mechanisms of soil.

The compression behaviour of agricultural soils can be described by three stages following the increase of vertical stress (Fig. 8):

(i) in Zone I where the vertical stresses are lower than the precompression stress, the volume change corresponds to the elastic deformation of the soil structure and the slope -d$e$/dlog$\sigma_v$ remains small;

(ii) (ii) in Zone II where the vertical stresses exceed the precompression stress, the volume change corresponds mainly to the reduction of the volume of the air-filled pores;

(iii) (iii) in Zone III, at high stresses, the volume of the air-filled pores becomes small and the reduction of meso-pores storing water dominates.

Note that all the compression curves are S-shaped in a sufficiently large stress range because the void ratio cannot become negative (Baumgartl and Köck (2004) proposed a minimum void ratio equal to 0.27 for the modelling). The shape of the compression curve depends on the part of the full-range curve involved by the applied stresses range (25 – 800 kPa, in the present work). If the stress applied is high enough to observe the convex part of the compression curve (Zone III in Fig. 8), the S-shape compression curve can be obtained.

Mitchell (1994) distinguished two main types of soil water: free water and adsorbed water; the adsorbed water that is strongly attracted to the clay surfaces has a structure different from that of free water. Because of the particularity of adsorbed water, it cannot be expelled out from the soil by the stresses as high as that considered in this study. As regards the free water, most of it is stored in meso-pores. This water is subjected to capillary forces as a function of the pores sizes. To expel the free water in meso-pores, high stresses are in general needed. As a result, when the decrease of the air-pores dominates during compression, the slope -d$e$/dlog$\sigma_v$ only depends on the soil structure because air (to a certain extent, also some water contained in the macro-pores) can flow easily. By contrast, when the decrease of meso-pores storing free water subjected to high capillarity dominates, the slope -d$e$/dlog$\sigma_v$ becomes smaller than the previous case because additional stresses are required to balance the capillary forces. Therefore, the compression curve has the bi-linear form when only the elastic deformation of the soil structure (Zone I) and the reduction of the air-pore volume (Zone II) occur. The S-shaped form appears when the flow of water subjected to high capillary forces in Zone III becomes significant.

This description of the mechanisms involved in the soil compression under vertical stresses is supported by different results in the literature about the evolution of the air phase and water phase under compression. Kutilek et al. (2006) analysed the pore size distribution of various soils and observed that the compression induced a significant decrease of the pore size in the structural domain (meso-pores). The changes in meso- and macro-porosity due to compression have been observed through a decrease of saturated hydraulic conductivity (Gebhardt et al., 2006; Dexter et al., 2004; Zhang et al., 2006) or a decrease of air permeability (Horn et al., 1995). In addition, Wiermann et al. (2000) and Schaffer et al. (2007) have observed the 3-D images of soil structure by X-ray tomography and found that the compaction of soil decreased both the porosity and the connectivity of the macro-pores. After Perdok et al. (2002) and Horn (2003), the reduction of the macro-pores affects first the air fraction and ultimately the water component.

The decrease of hydraulic conductivity during compaction is also an important factor affecting the compaction curve. In the adopted experimental protocol, the stress applied on the soil samples was increased every 5 min. As the hydraulic conductivity decreases when the stress is increasing, it is possible that the water in the soil was not fully expelled within this duration of 5 min. This could also explain the S-shaped compression curve observed from some tests; further investigations (for example, loading at various durations) are required to clarify this point.

As far as the effect of clay content is concerned, the compression tests performed on the remould soils showed that the S-shaped curves were obtained more frequently on the soil having higher clay content. For instance, almost all the tests of soil D (0.47 of clay content) had the S-shaped curve and almost all the tests of soil A (0.19 of clay content) had the bi-linear compression curve. This could be also explained by the effect of capillary forces.

With a high clay content, high capillary forces can be expected in the soil due to the air-water interactions in the meso-pores. In consequence, as explained previously, the difference between the slope -d$e$/dlog$\sigma_v$ in the Zone II and that in the Zone III must be more pronounced at a higher clay content. In a low plasticity soil, as lower capillary forces can be expected in the meso-pores, the slope -d$e$/dlog$\sigma_v$ obtained when decreasing the air-filled pores (Zone II) and that when decreasing the meso-pores storing free water (Zone III) is similar. As a consequence, the bi-linear compression curve can be observed.

The effect of soil moisture evidenced by the results presented in Fig. 3 from the compression tests on the remould specimens of soil C (0.34 of clay content) can be interpreted following the schema of Fig. 8. For the tests having low $S_{ri}$ or high initial $V_a/V$ (T18, T21, T24; $w_i$ equal to 17%), the compression curve was dominated by the decrease of air-filled pores and thus the slope -d$e$/dlog$\sigma_v$ remained constant from the precompression stress until the highest stress (800 kPa). At the highest stress, $V_a/V$ remained higher than 0.1. For the tests having high $S_{ri}$ or low initial $V_a/V$ (T20, T23, T26; $w_i$ equal to 28.5%), the decrease of the meso-pores storing water that subjected to high capillary forces dominated during all the compression process. Thus, the slope -d$e$/dlog$\sigma_v$ remained constant during all the compression stage as its value was smaller than that observed in T18, T21, T24 ($w_i$ equal to 17%) that were described previously (see the results of $C_c$ in Fig. 7). Because the initial stress of the compression curve (15 kPa) was higher than the precompression stress in T20, T23, and T26, only Zone II and Zone III were observed from this test. Finally, for T19, T22, T25 ($w_i$ equal to 22%), two slopes were observed on the compression curve under the stresses higher than the precompression stress. The first slope corresponds to the compression of air-filled pores and its values were similar to that of the group having $w_i$ equal to 17% (see the results of $C_c$ in Fig. 7). The second slope corresponds to the phase where the decrease of the meso-pores storing water that were subjected to high capillary forces dominated; its values were similar to that of the group having $w_i$ equal to 28.5% (see the plot -d$e$/dlog$\sigma_v$ versus log$\sigma_v$ in Fig. 3). All the three zones were observed in T19, T22, and T25.

The S-shaped compression curves were observed more frequently on the remould soils than on the undisturbed soil. This difference can be explained by the difference in soil structure. Wiermann et al. (2000) analysed the computer tomographic images and observed a difference in the configuration of the pore space between ploughed soil and untilled soil. The ploughed soil had an irregular arrangement of connected inter-aggregate packing pores (meso-pores) while the untilled soil had channels with preferred vertical orientation embedded in a compact soil matrix (macro-pores). In addition, Wiermann et al. (2000) noted that these channels, which had been formed by roots and worms, were preserved under loading for the soil at 30-40 cm depth. In consequence, the compression behaviour of undisturbed soil is strongly dependent on the presence of these macro-pores. That explains also the insignificance of the effect of $w_i$ on $C_c$ (see Fig. 7$d$).

The effect of $\rho_i$ on $C_c$ observed on the remould soils in the present work is in agreement with that observed previously by Imhoff et al. (2004): $C_c$ decreased with the increase of $\rho_i$ (Fig. 7$a, b, c$). Indeed, $C_c$ represents the compressibility of soil and the soil is more compressible at lower dry density.

The effect of $w_i$ on $C_c$ is strongly depending on the soil properties. In the case of soil A, $w_i$ had no effect on $C_c$ (Fig. 7*a*), that is in agreement with the results of Imhoff et al. (2004). The effect of $w_i$ on $C_c$ observed on soil D is in contradiction with that generally observed on unsaturated soils: Cui and Delage (1996) observed that $C_c$ increased with decreasing matric suction (increase of $w_i$), while the tests on soil D showed an increase of $C_c$ when $w_i$ decreased (Fig. 7*c*). Pereira et al. (2007) performed compression tests on a silty soil (silt content equal to 0.67) and found that $C_c$ increased with $w_i$ increase in the range of low $w_i$, and $C_c$ decreased with $w_i$ increase in the range of high $w_i$. In general, when $w_i$ is low, the decrease of air-filled pores dominates during the compression, thus $C_c$ mainly depends on the soil structure. The increase of $w_i$ reduces the matric suction, softens the soil aggregates as well as the bond between them, leading to a $C_c$ increase. By contrast, when $w_i$ is high, the decrease of the meso-pores storing water dominates during the compression. When the water subjected to capillary forces is involved, $C_c$ depends not only on the soil structure but also on the capillarity, i.e., the higher the water content, the higher the stress required to expel the capillary water during the compression, thus the lower is compression index $C_c$. In the case of the soil at low plasticity (soil A), the capillarity is insignificant and the structure of the aggregates is also not sensitive to the change of water content. For this reason, $C_c$ is independent on $w_i$. On the contrary, for the soil D of high plasticity (clay), the pores were smaller and thus high capillarity can be expected in it. The effect of $w_i$ on $C_c$ was then significant. Regarding the undisturbed soils (Fig. 7*d*), as the soil mechanical properties are strongly influenced by its structure (Horn and Lebert, 1994), the effect of $w_i$ was not significant.

Note that in the present work, the description of the soils pore structure and the mechanisms of water retention have been used to interpret the mechanical behaviour of the soils. Further investigations on the soils structure are needed to highlight the link between the mechanical loading and the changes in soils structure.

## 5. *Conclusion*

The compression curve of agricultural soils obtained from confined uniaxial tests has usually two main forms: bi-linear and S-shaped. The choice of constitutive model when determining the soil compaction is then strongly depending on the form of the compression curve. Experimental results on the remould soils showed that the S-shaped curves were observed more frequently when the soil plasticity and/or the initial water content are high. In the present work, within the range of vertical stress of 15 – 800 kPa, the S-shaped curves were obtained on all the remould samples of soil D (clay) and on some remould samples of soil C (silty clay) but on none of remould samples of soil A (sandy loam). With the undisturbed soils, the S-shaped curves were observed less frequently than with the remould soils. Note however that as this experiment did not include replicates, more results are required in order to generalize these conclusions.

A new approach for analyzing the compression curve is proposed. It consists in dividing the curve in three zones: Zone I, at stresses lower than the precompression stress, corresponds to the elastic behaviour of the soil; Zone II, when the stresses exceeds the precompression stress and the collapse of air-filled pores occurs; Zone III, at higher stresses, when the volume of air-filled pores becomes small and the mechanism of capillary water flow becomes dominant. When water flow is involved, increasing stress decreases the volume and meanwhile redistributes the soil water. The difference between the slope -

d$e$/dlog$\sigma_v$ in Zone III and that in Zone II was found to be more significant at higher water content.

## 6. Acknowledgements

This work was carried out under the project "Soil degradation due to compaction" with the financial support of (1) the « ANR- Agence Nationale de la Recherche - The French National Research Agency » under the « Programme Agriculture et Développement Durable », project « ANR-05-PADD-013 », (2) the Ministry in charge of Environment under the programme GESSOL2 "Impact des pratiques agricoles sur le sol et les eaux ». The authors are also grateful for the technical assistance of D. Boitez and F. Bornet.

**Table 1. Some physical and chemical properties of the studied soils.**

| Soil | A | B | C | D |
|---|---|---|---|---|
| Site | Breuil | Mons | Avignon | Epernay |
| Particle density (Mg/m$^3$) | 2.56 | 2.62 | 2.71 | 2.68 |
| Liquid limit (%) | 58 | 32 | 31 | 49 |
| Plastic limit (%) | 51 | 22 | 20 | 29 |
| Plasticity index (%) | 7 | 10 | 11 | 20 |
| Organic carbon content (g/kg) | 82.8 | 8.5 | 10.2 | 16.8 |
| Methylene blue absorption (g/100g) | 0.4 | 1.4 | 2.3 | 7.4 |
| Particle size distribution (g/g): | | | | |
|     Clay (< 2 µm) | 0.19 | 0.19 | 0.34 | 0.47 |
|     Silt (2 – 50 µm) | 0.23 | 0.75 | 0.51 | 0.33 |
|     Sand (50 – 2000 µm) | 0.58 | 0.06 | 0.16 | 0.20 |
| USDA Classification | Sandy loam | Silt loam | Silty clay loam | Clay |
| FAO Classification | Medium | Medium | Medium fine | Fine |
| FAO Taxonomy | Dystric cambisol | Hortic luvisol | Calcaric cambisol | Calcaric cambisol |

**Table 2. Summary of the compression tests, initial and final soil properties.**

| No. | Soil | Classification | Layer (cm) | $e_i$ | $e_f$ | $w_i$ (%) | $w_f$ (%) | $\rho_i$ (Mg/m³) | $\rho_f$ (Mg/m³) | $S_{ri}$ | $S_{rf}$ |
|---|---|---|---|---|---|---|---|---|---|---|---|
| 1 | A | Medium | 0-30 | 1.43 | 0.81 | 18.9 | 18.2 | 1.05 | 1.41 | 0.34 | 0.58 |
| 2 | A | Medium | 0-30 | 1.49 | 0.85 | 24.7 | 24.7 | 1.03 | 1.38 | 0.42 | 0.74 |
| 3 | A | Medium | 0-30 | 1.49 | 0.81 | 25.3 | 22.7 | 1.03 | 1.41 | 0.43 | 0.72 |
| 4 | A | Medium | 0-30 | 1.09 | 0.80 | 17.9 | 18.8 | 1.22 | 1.42 | 0.42 | 0.60 |
| 5 | A | Medium | 0-30 | 1.08 | 0.73 | 23.3 | 22.5 | 1.23 | 1.48 | 0.55 | 0.79 |
| 6 | A | Medium | 0-30 | 1.16 | 0.79 | 24.7 | 24.7 | 1.19 | 1.43 | 0.55 | 0.80 |
| 7 | A | Medium | 0-30 | 0.93 | 0.76 | 20.2 | 19.1 | 1.33 | 1.45 | 0.56 | 0.64 |
| 8 | A | Medium | 0-30 | 0.99 | 0.77 | 24.2 | 24.2 | 1.29 | 1.45 | 0.63 | 0.80 |
| 9 | A | Medium | 0-30 | 1.01 | 0.74 | 25.8 | 25.2 | 1.27 | 1.47 | 0.65 | 0.87 |
| 10 | A | Medium | 30-60 | 1.30 | 0.84 | 16.3 | 13.5 | 1.11 | 1.39 | 0.32 | 0.41 |
| 11 | A | Medium | 30-60 | 1.41 | 0.83 | 18.1 | 17.4 | 1.06 | 1.40 | 0.33 | 0.54 |
| 12 | A | Medium | 30-60 | 1.24 | 0.72 | 24.7 | 23.8 | 1.14 | 1.49 | 0.51 | 0.85 |
| 13 | B | Medium | 30-60 | 0.59 | 0.51 | 12.4 | 12.4 | 1.65 | 1.70 | 0.55 | 0.62 |
| 14 | B | Medium | 30-60 | 0.75 | 0.62 | 18.2 | 18.2 | 1.50 | 1.58 | 0.64 | 0.75 |
| 15 | B | Medium | 30-60 | 0.79 | 0.63 | 19.5 | 19.5 | 1.46 | 1.57 | 0.65 | 0.79 |
| 16 | B | Medium | 30-60 | 0.75 | 0.57 | 27.1 | 19.9 | 1.50 | 1.63 | 0.95 | 0.89 |
| 17 | B | Medium | 30-60 | 0.78 | 0.60 | 29.7 | 20.5 | 1.47 | 1.60 | 1.00 | 0.87 |
| 18 | C | Medium fine | 0-30 | 1.36 | 0.66 | 16.8 | 16.4 | 1.15 | 1.54 | 0.34 | 0.64 |
| 19 | C | Medium fine | 0-30 | 1.24 | 0.59 | 22.4 | 19.5 | 1.21 | 1.61 | 0.49 | 0.85 |
| 20 | C | Medium fine | 0-30 | 1.04 | 0.64 | 28.7 | 20.4 | 1.33 | 1.56 | 0.75 | 0.82 |
| 21 | C | Medium fine | 0-30 | 1.13 | 0.63 | 16.5 | 16.2 | 1.27 | 1.57 | 0.40 | 0.66 |
| 22 | C | Medium fine | 0-30 | 1.10 | 0.58 | 22.1 | 19.5 | 1.29 | 1.62 | 0.54 | 0.86 |
| 23 | C | Medium fine | 0-30 | 0.91 | 0.60 | 28.5 | 20.2 | 1.42 | 1.60 | 0.85 | 0.86 |
| 24 | C | Medium fine | 0-30 | 1.06 | 0.64 | 16.8 | 16.4 | 1.32 | 1.56 | 0.43 | 0.66 |
| 25 | C | Medium fine | 0-30 | 0.98 | 0.59 | 21.9 | 19.6 | 1.37 | 1.61 | 0.60 | 0.85 |
| 26 | C | Medium fine | 0-30 | 0.88 | 0.63 | 28.2 | 19.6 | 1.44 | 1.57 | 0.87 | 0.80 |
| 27 | C | Medium fine | 30-60 | 0.76 | 0.60 | 19.2 | 18.4 | 1.54 | 1.60 | 0.68 | 0.79 |
| 28 | C | Medium fine | 30-60 | 0.79 | 0.61 | 21.0 | 19.8 | 1.51 | 1.59 | 0.72 | 0.83 |
| 29 | C | Medium fine | 30-60 | 0.80 | 0.60 | 23.0 | 19.2 | 1.51 | 1.60 | 0.78 | 0.82 |
| 30 | D | Fine | 0-30 | 1.74 | 0.82 | 25.6 | 25.2 | 0.98 | 1.41 | 0.39 | 0.79 |
| 31 | D | Fine | 0-30 | 1.64 | 0.84 | 32.5 | 28.6 | 1.02 | 1.39 | 0.53 | 0.87 |
| 32 | D | Fine | 0-30 | 1.44 | 0.87 | 37.8 | 30.4 | 1.10 | 1.37 | 0.70 | 0.89 |
| 33 | D | Fine | 0-30 | 1.41 | 0.78 | 25.6 | 25.0 | 1.11 | 1.44 | 0.49 | 0.82 |
| 34 | D | Fine | 0-30 | 1.36 | 0.86 | 31.3 | 28.7 | 1.14 | 1.38 | 0.62 | 0.85 |
| 35 | D | Fine | 0-30 | 1.27 | 0.88 | 37.1 | 29.7 | 1.18 | 1.36 | 0.78 | 0.86 |
| 36 | D | Fine | 0-30 | 1.11 | 0.79 | 25.1 | 24.2 | 1.27 | 1.43 | 0.60 | 0.78 |
| 37 | D | Fine | 0-30 | 1.10 | 0.82 | 30.4 | 28.4 | 1.28 | 1.41 | 0.74 | 0.89 |
| 38 | D | Fine | 0-30 | 1.17 | 0.90 | 37.9 | 31.5 | 1.24 | 1.35 | 0.87 | 0.90 |
| 39 | D | Fine | 30-60 | 1.08 | 0.90 | 30.2 | 29.6 | 1.29 | 1.35 | 0.75 | 0.84 |
| 40 | D | Fine | 30-60 | 1.16 | 0.90 | 32.5 | 30.6 | 1.24 | 1.35 | 0.75 | 0.87 |
| 41 | D | Fine | 30-60 | 1.49 | 1.15 | 41.4 | 39.6 | 1.08 | 1.19 | 0.74 | 0.88 |

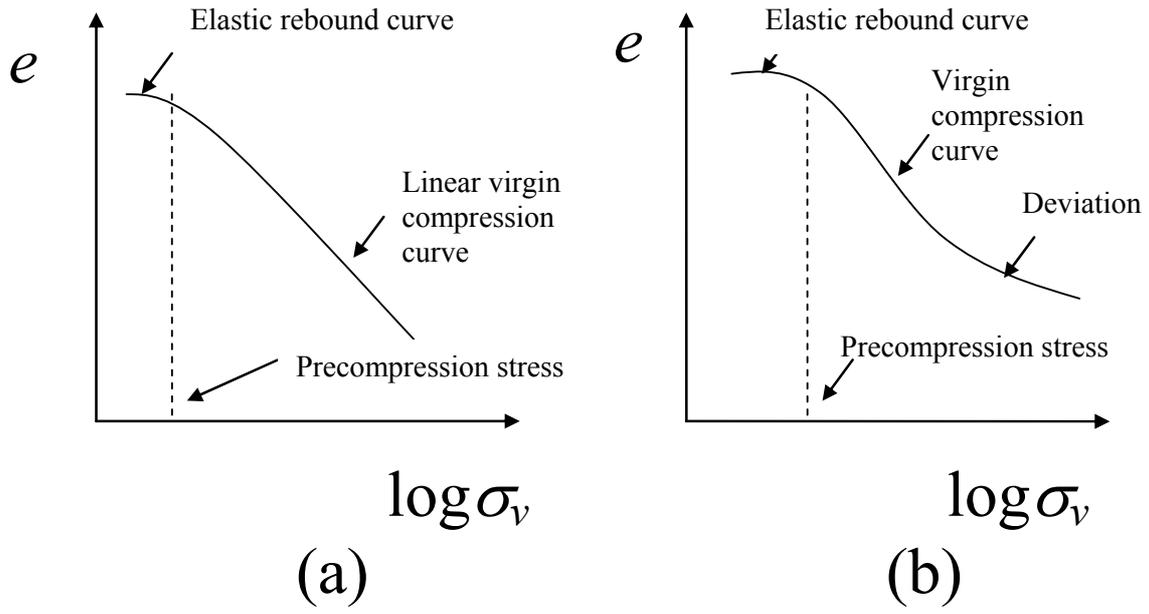

**Fig. 1. Two main forms of confined uniaxial compression curve: (A) bi-linear curve; (B) S-shaped curve.**

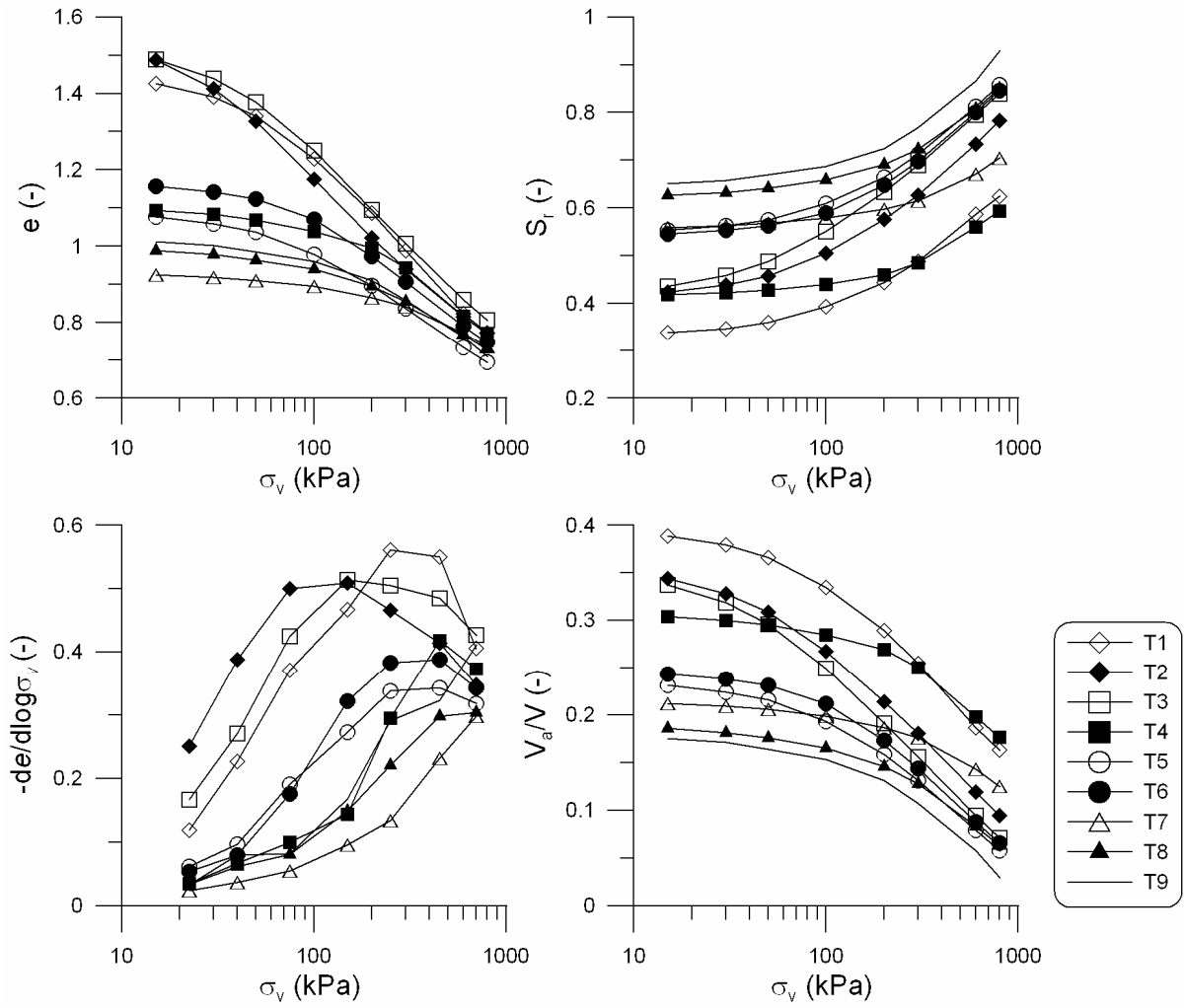

**Fig. 2.** Compression tests of soil A for 0 – 30 cm layer. Void ratio ($e$), degree of saturation ($S_r$), - d$e$/dlog$\sigma_v$ and air content ($V_a/V$) versus logarithm of vertical stress ($\sigma_v$).

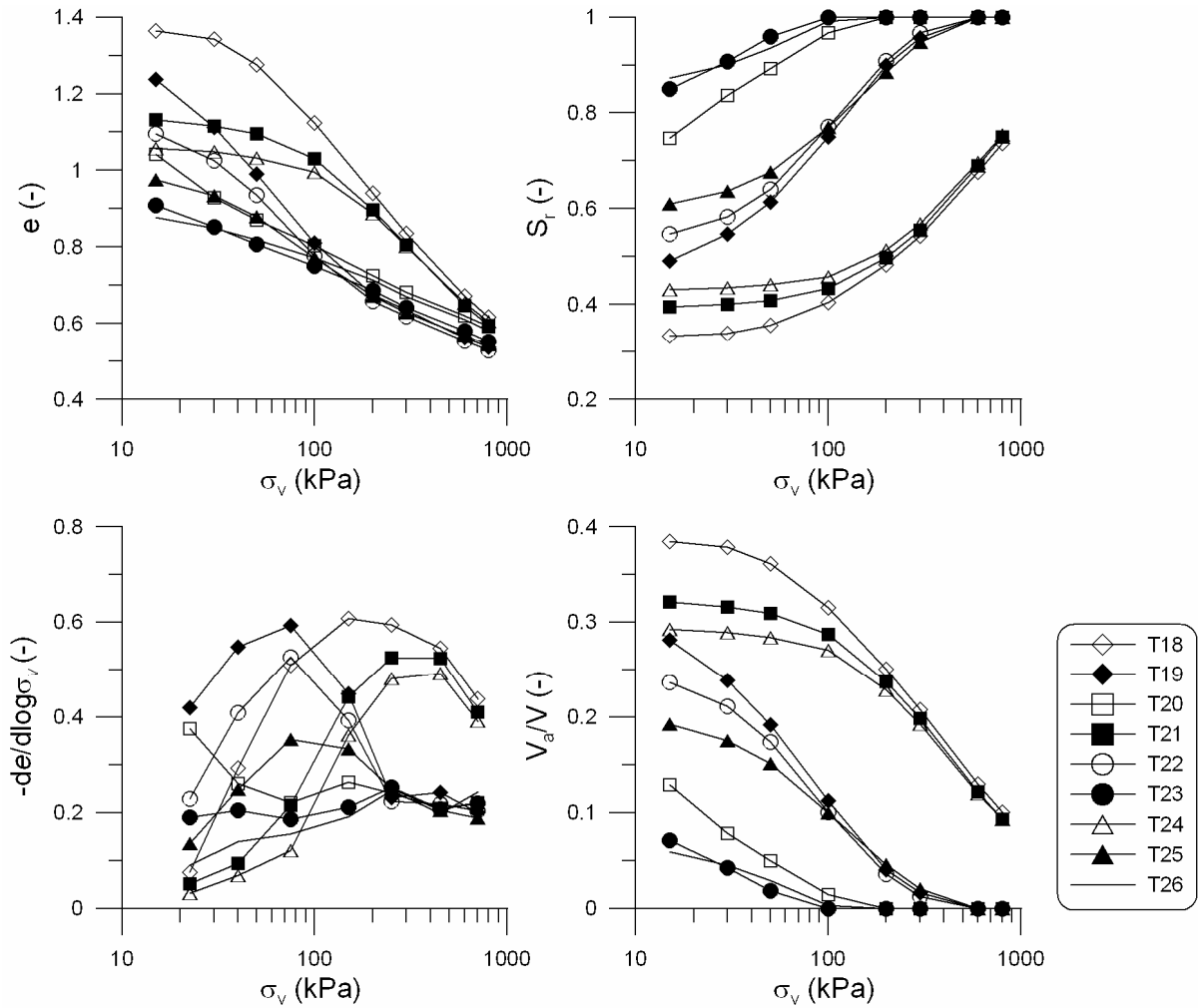

Fig. 3. Compression tests of soil C for 0 – 30 cm layer. Void ratio (*e*), degree of saturation (*$S_r$*), - d*e*/dlog$\sigma_v$ and air content (*$V_a/V$*) versus logarithm of vertical stress (*$\sigma_v$*).

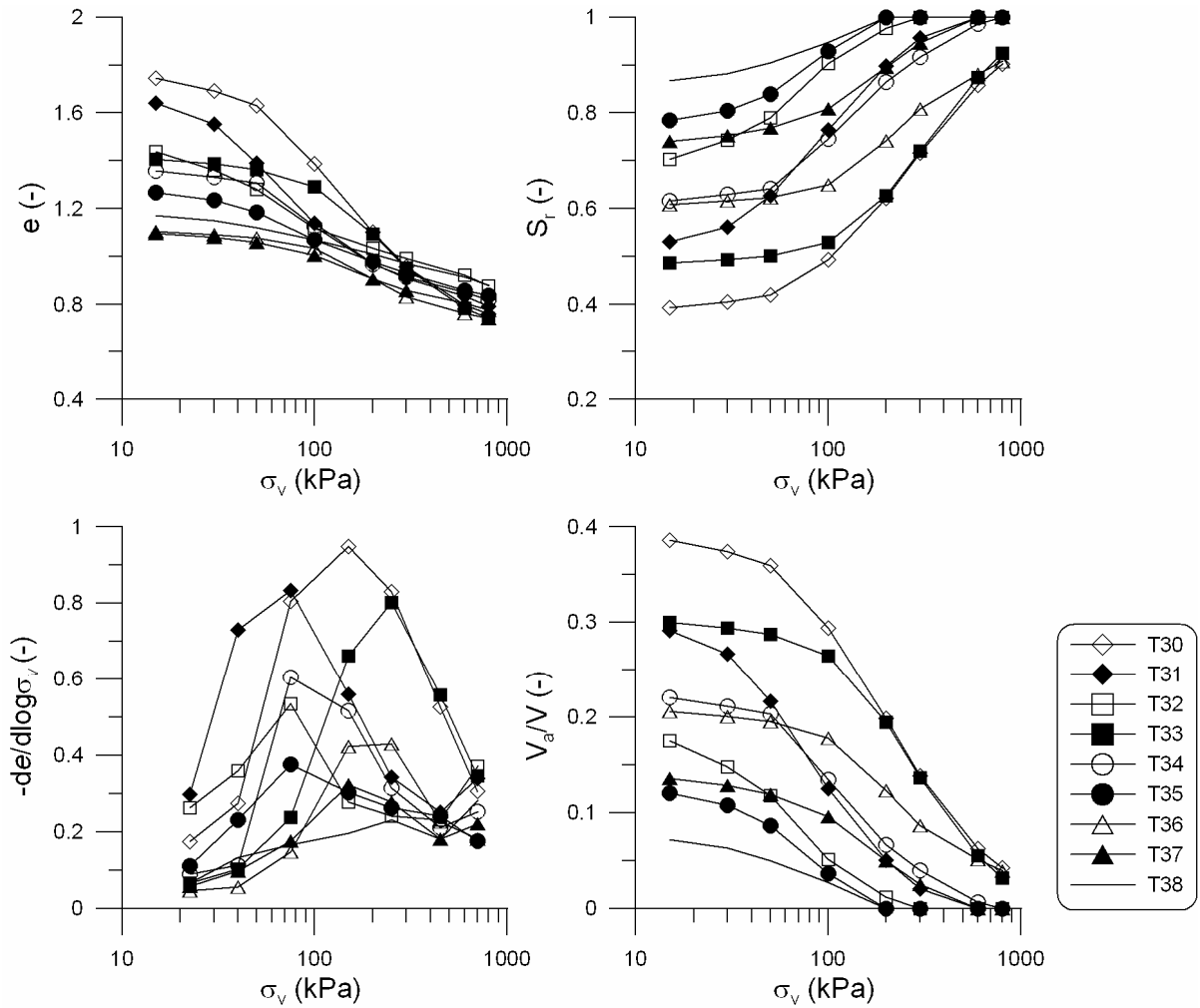

Fig. 4. Compression tests of soil D for 0 – 30 cm layer. Void ratio (*e*), degree of saturation ($S_r$), - d*e*/dlog$\sigma_v$ and air content ($V_a/V$) versus logarithm of vertical stress ($\sigma_v$).

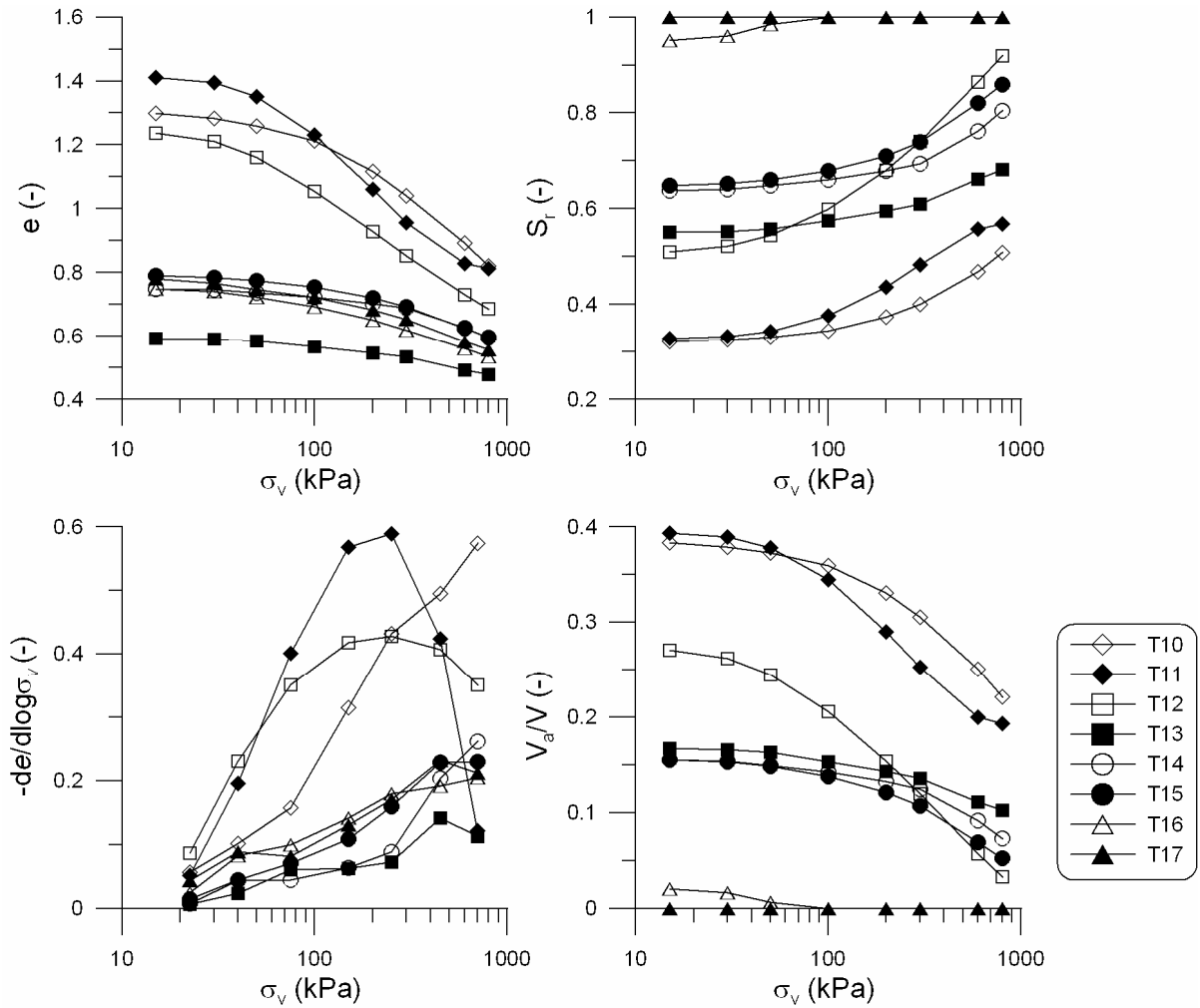

Fig. 5. Compression tests of soils A and B for 30 – 60 cm layer. Void ratio (*e*), degree of saturation ($S_r$), - d*e*/dlog$\sigma_v$ and air content ($V_a/V$) versus logarithm of vertical stress ($\sigma_v$).

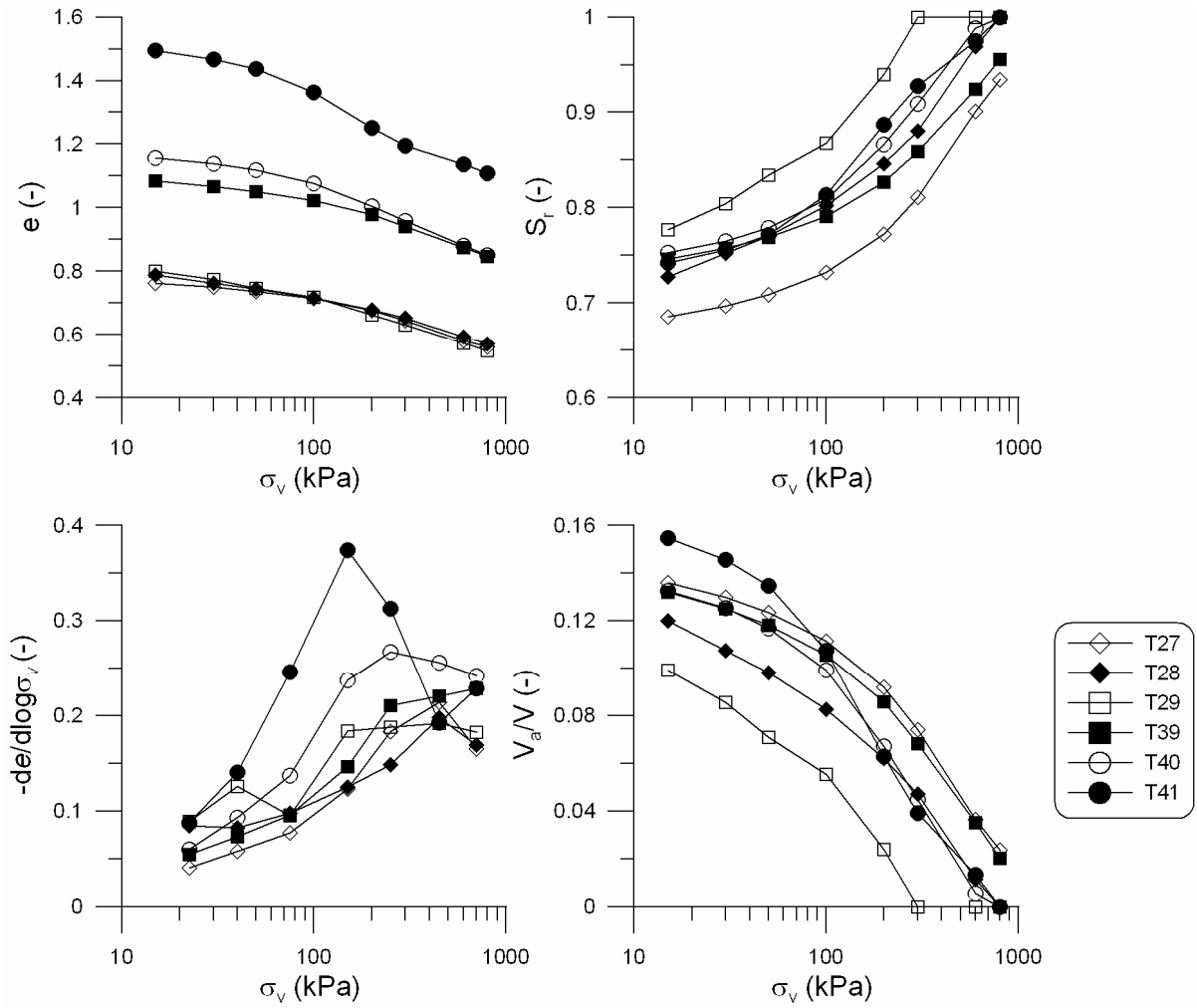

**Fig. 6. Compression tests of soils C and D for 30 – 60 cm layer. Void ratio (*e*), degree of saturation (*S*$_r$), - d*e*/dlog$\sigma_v$ and air content (*V*$_a$/*V*) versus logarithm of vertical stress ($\sigma_v$).**

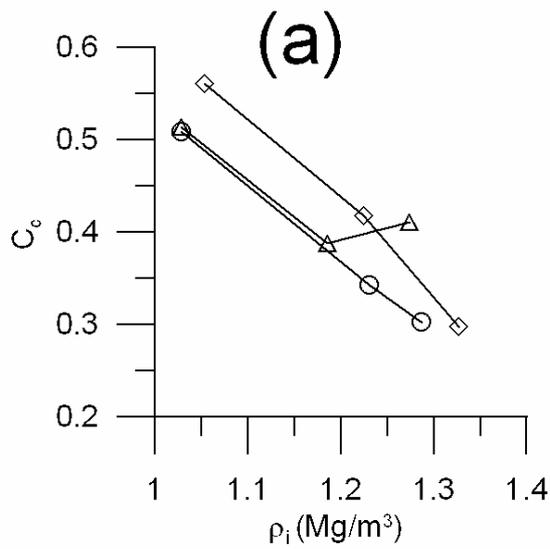
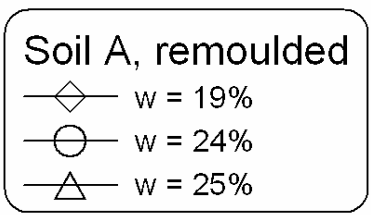
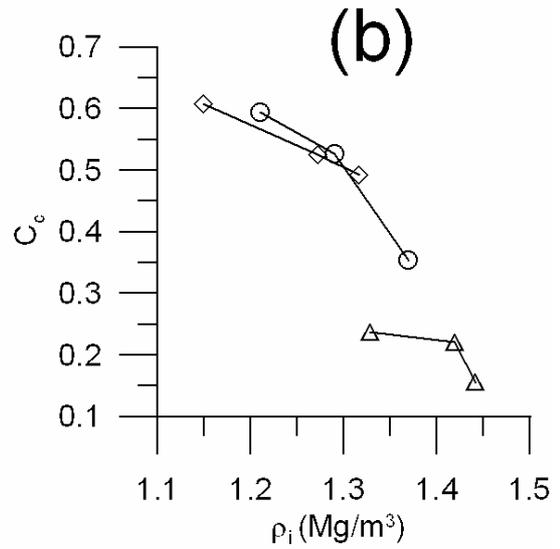
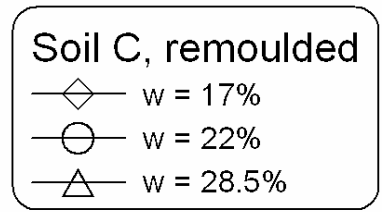
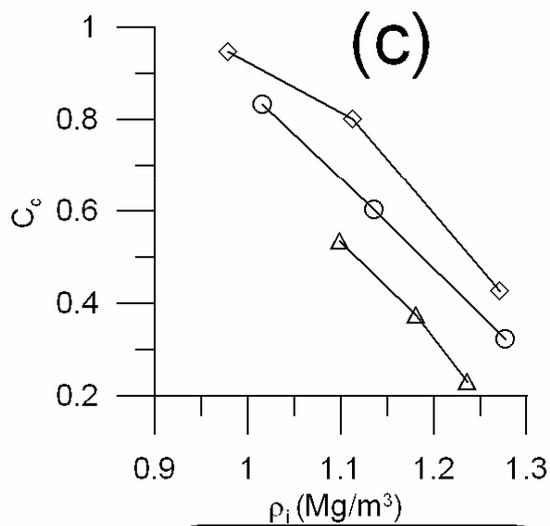
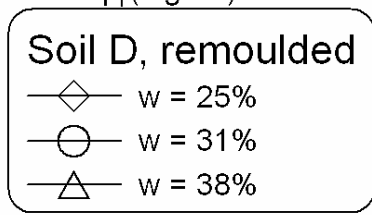
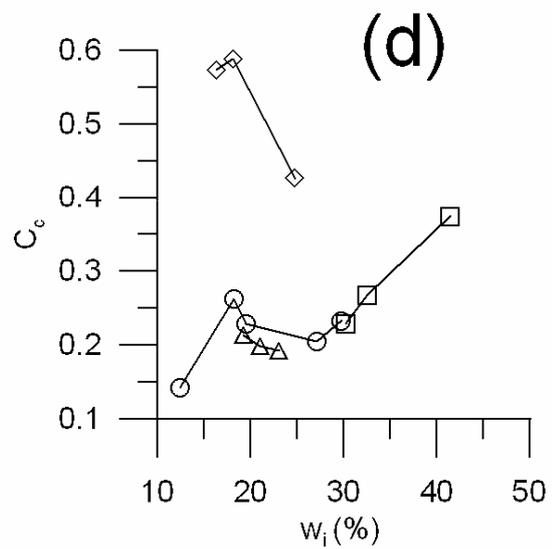
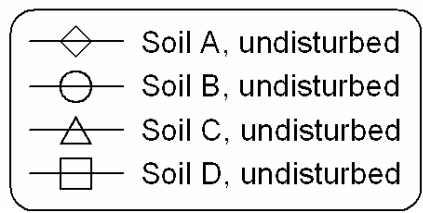

Fig. 7. Compression index ($C_c$) at various initial dry bulk densities ($\rho_i$) and initial water contents ($w_i$).

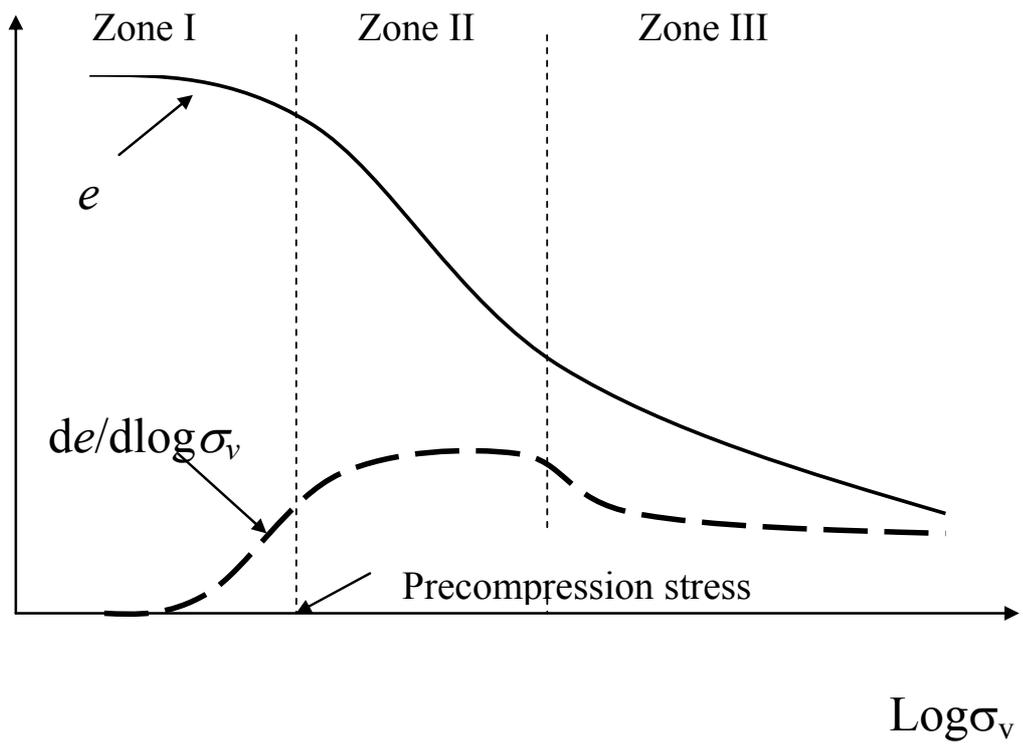

**Fig. 8. Complete representative compression curve.**